\documentclass[prb,twocolumn]{revtex4}%

\usepackage{graphicx}
\usepackage{amsmath}
\usepackage{ulem}
\usepackage{color}

\newcommand{\be}{\begin{equation}}
\newcommand{\ee}{\end{equation}}
\newcommand{\bea}{\begin{eqnarray*}}
\newcommand{\eea}{\end{eqnarray*}}
\newcommand{\bean}{\begin{eqnarray}}
\newcommand{\eean}{\end{eqnarray}}

\begin{document}

\draft
\title
{\bf Tunneling Current Spectra of a Metal Core/Semiconductor Shell
Quantum Dot Molecule}

\author{David M.-T. Kuo}
\address{Department of Electrical Engineering, National Central
University, Chungli 320, Taiwan}



\begin{abstract}
The transport properties of a metal core/semiconductor shell quantum
dot molecule (QDM) embedded in a matrix connected to metallic
electrodes are theoretically studied in the framework of Keldysh
Green function technique. The effects of the electron plasmon
interactions (EPIs) on the tunneling current spectra of QDM are
examined. The energy levels of the QDs, intradot and interdot
Coulomb interactions, electron interdot hopping strengths, and
tunneling rates of QDs are renormalized by the EPIs. The
differential conductance spectra show peaks arising from the plasmon
assisted tunneling process, intradot and interdot Coulomb
interactions, and coherent tunneling between the QDs.
\end{abstract}


\maketitle


\section{Introduction}

Although many of the interesting transport phenomena related to
quantum dots (QDs) including the Coulomb blockade,$^{1)}$ Kondo
effect,$^{2)}$ Fano resonance,$^{3)}$ Pauli spin blockade,$^{4)}$
photon (phonon) assisted tunneling process$^{5)}$ and negative
differential conductance,$^{6)}$ have already been extensively
studied, there has been little reported on the effects of electron
plasmon interactions (EPIs) on the tunneling current of QDs.
Recently, EPIs have received considerable attention for their
applications in nanophotonics, biology, and the harvesting of solar
energy.$^{7)}$ Much effort has been focused on the effects of EPIs
on the optical properties of nanostructures.$^{8-11)}$ Experiments
can observe the strong effect of EPIs on the exciton spectrum of
individual semiconductor QDs arising from adjunctive metallic
nanostructures$^{12,13)}$.

In the application of QD-based biosensor, the metal
core/semiconductor shell QDs play an crucial role.$^{14)}$ Because
this type QDs can be readily coupled to detected proteins through
electrostatic interactions. Despite the measurement of optical
spectra can resolve molecules,$^{7-14)}$ tunneling current spectra
provide a high efficient means to identify molecules and address
electrically a single nano-object.$^{15,16)}$ This inspires us to
study the tunneling current through individual metal
core/semiconductor shell QDs for the application of nanoscale
biosensor. However, it is difficult to avoid the proximity effect
between such type QDs since QDs are randomly distributed.$^{17)}$
Therefore, we consider a metal core/semiconductor shell QD molecule
(QDM) embedded in a matrix connected to metallic electrodes shown in
Fig. 1 to clarify how the EPIs to influence the proximity effect
resulting from interdot electron Coulomb interactions, electron
hopping process, and plasmon hopping between two QDs in the absence
of detected proteins.

\section{Formalism}

Here, we consider nanoscale semiconductor QDs, the energy levels
separation of each QD is much larger than their on-site Coulomb
interactions and thermal energies. One energy level for each quantum
dot is considered in this study. The two-level Anderson model
including EPIs is employed to simulate the QDM junction system as
shown in Fig. 1. The Hamiltonian of the QDM junction is given by
$H=H_{0}+H_{QDM}+H_T$

\begin{eqnarray}
H_0& = &\sum_{k,\sigma} \epsilon_k
a^{\dagger}_{k,\sigma}a_{k,\sigma}+ \sum_{k,\sigma} \epsilon_k
b^{\dagger}_{k,\sigma}b_{k,\sigma}\\ \nonumber
&+&\sum_{k,\ell,\sigma}
V_{k,\ell,L}d^{\dagger}_{\ell,\sigma}a_{k,\sigma}
+\sum_{k,\ell,\sigma}V_{k,\ell,R}d^{\dagger}_{\ell,\sigma}b_{k,\sigma}+c.c
\end{eqnarray}
where the first two terms describe the free electron gas at the left
and right metallic electrodes. $a^{\dagger}_{k,\sigma}$
($b^{\dagger}_{k,\sigma}$) creates  an electron of momentum $k$ and
spin $\sigma$ with energy $\epsilon_k$ at the left (right) metallic
electrode. $V_{k,\ell,\beta}$ ($\ell=1,2$) describes the coupling
between the metallic electrodes and the QDs.
$d^{\dagger}_{\ell,\sigma}$ ($d_{\ell,\sigma}$) creates (destroys)
an electron in the $\ell$-th dot:

\begin{small}
\begin{eqnarray}
H_{QDM}&=& \sum_{\ell,\sigma} E_{\ell} n_{\ell,\sigma}+
\sum_{\ell} U_{\ell} n_{\ell,\sigma} n_{\ell,\bar\sigma}\\
\nonumber &+&\frac{1}{2}\sum_{\ell,j,\sigma,\sigma'}
U_{\ell,j}n_{\ell,\sigma}n_{j,\sigma'}
+\sum_{\ell,j,\sigma}t_{\ell,j} d^{\dagger}_{\ell,\sigma}
d_{j,\sigma},
\end{eqnarray}
\end{small}
where $E_{\ell}$ is the spin-independent QD energy level, and
$n_{\ell,\sigma}=d^{\dagger}_{\ell,\sigma}d_{\ell,\sigma}$.
Notations $U_{\ell}$ and $U_{\ell,j}$ describe the intradot and
interdot Coulomb interactions, respectively. $t_{\ell,j}$ describes
the electron interdot hopping. The Hamiltonian of QD molecule
described by Eqs. (1) and (2) has already been extensively
considered for studying the transport properties of
nanostructures.$^{18-20)}$ The Hamiltonian of electron plasmon
interactions (EPIs) arising from the metallic nanostructure of each
QDs can be written as $H_T$:
\begin{equation}
H_T=  \sum_{\ell=1,2} \omega_{\ell}c^{\dagger}_{\ell} c_{\ell}+
\sum_{\ell,\sigma} \Omega_{\ell} n_{\ell,\sigma}
(c^{\dagger}_{\ell}+c_{\ell})-\sum_{i \neq \ell} \omega_{\ell,j}
c^{\dagger}_{\ell} c_{j},
\end{equation}
where $\omega_{\ell}$ is the plasmon frequency of the metallic
nanostructures shown in Fig. 1, and $\Omega_{\ell}$ is the coupling
strength of EPIs. The last term involving $\omega_{\ell,j}$
describes the plasmon hopping between two metallic nanostructures.
When metal nanostructures are close enough, the plasmonic modes on
one nanostructure can couple with that on other nanostructures so
that hopping of plasmons from one metallic nanostructure to other
nanostructure becomes possible.

A canonical transformation can be used to remove the on-site EPIs
from Eq. (3), that is $H_{new}= e^{S^{\dagger}} H e^S$, where
$S=-\sum_{\ell,\sigma}\Omega_{\ell} n_{\ell,\sigma}
(c^{\dagger}_{\ell}-c_{\ell})$.$^{21)}$ In the new Hamiltonian, we
have the following effective physical parameters:
$V^e_{k,1.\alpha}=V_{k,1,\alpha}e^{\lambda_1(c^{\dagger}_1-c_1)}$,
$V^e_{k,2,\alpha}=V_{k,2,\alpha}e^{\lambda_2 (c^{\dagger}_2-c_2)}$,
$E^e_{\ell}=E_{\ell}-\lambda^2_{\ell}\omega_{\ell}$,
$U^e_{\ell}=U_{\ell}-2\lambda^2_{\ell} \omega_{\ell}$,
$U^e_{\ell,j}=U_{\ell,j}-2\lambda_1 \lambda_2 \omega_{12}$,
$t^e_{\ell,j}=t_{\ell,j}
e^{-[\lambda_{1}(c^{\dagger}_{1}-c_1)-\lambda_2(c^{\dagger}_2-c_2)]}$,
and $\lambda_{\ell}=\Omega_{\ell}/\omega_{\ell}$. In addition,
$H_{new}$ has an extra term $\sum_{\ell \neq j,\sigma}
\lambda_{\ell} \omega_{\ell,j}n_{\ell,\sigma}(c^{\dagger}_{j}+c_j)$
resulting from the plasmon hopping between the metal nanostructures,
which vanishes under free plasmon average introduced later. Under
canonical transformation, we note that the narrowing of QD energy
level broadening appears and QD energy levels, electron Coulomb
interactions, and electron interdot hopping strengths become
renormalized as well. The proximity effect between QDs resulting
from $U^e_{\ell,j}$ and $t^e_{\ell,j}$ is influenced by the EPIs. It
is possible to observe an effective negative interdot Coulomb
interactions if we have strong EPIs and a large hopping matrix
element of plasmon ($\omega_{\ell,j}$). For simplicity, this study
is restricted in the case of $U^e_{\ell,j} \ge 0$ and identical QDs
($\omega_{\ell}=\omega_0$ and $\lambda_{\ell}=g$ ).

To decuple the EPIs of $H_{new}$, we take the mean-field average to
remove the plasmon field arising from $c^{\dagger}-c$, that is $<
e^{g(c^{\dagger}-c)} > =e^{-\frac{1}{2}g^2
coth^2(\hbar\omega_0/(2k_BT))}$. Based on such a mean-field average,
we see a reduction of $V_{k,\ell,\alpha}e^{-\frac{1}{2}g^2
coth^2(\hbar\omega_0/(2k_BT))}=V_{k,\ell,\alpha}X_{\ell}$ and
interdot hopping strength $t_c
e^{-\frac{1}{2}(\lambda_{1}-\lambda_2)^2
coth^2(\hbar\omega_0/(2k_BT))}=t_cX_{\ell,j}=t_c$. This
approximation is valid when the tunneling rate arising from the
coupling between the QDs and the electrodes is smaller than the
EPIs,$^{22)}$ which is our condition of interest. Based on such an
approximation, the $\omega_0$-dependent tunelling rates are
neglected in this study. Meanwhile, we assume that there is no
voltage difference between two dots. This implies that the tunneling
currents directly involving $t_c$ can be ignored.

Using the Keldysh-Green's function technique,$^{23)}$ the tunneling
currents of the QDMs in the Coulomb blockade regime are given by
\begin{equation}
J=\frac{-e}{h}\sum_{\ell,\sigma} \int d\epsilon
\frac{\Gamma_{\ell,L}(\epsilon)\Gamma_{\ell,R}(\epsilon)}{\Gamma_{\ell,L}(\epsilon)+\Gamma_{\ell,R}(\epsilon)}
ImG^r_{\ell,\sigma}(\epsilon) [f_L(\epsilon)-f_R(\epsilon)]
\end{equation}
where $f_{L(R)}(\epsilon)=1/[e^{(\epsilon-\mu_{L(R)})/k_BT}+1]$
denotes the Fermi distribution function of the left (right)
electrode. The left (right) chemical potential is given by
$\mu_L(\mu_R)$. $\mu_L-\mu_R=e V_a$, where $ V_a$ denotes the
applied bias. Notation $T$ denotes the equilibrium temperature of
the left and right electrodes. $e$ and $h$ denote the electron
charge and Planck's constant, respectively. Notation
$\Gamma_{\ell,L(R)}=\sum_{k}V^2_{k,\ell,L(R)} \delta
(\epsilon-\epsilon_k)$ denotes the tunnelling rate from the left
(right) electrode to dot 1 and dot 2, which is assumed to be energy-
and bias-independent. The retarded Green function of the QD density
of states ($-ImG^r_{\ell,\sigma}$) has the following expression
\begin{eqnarray}
G^r_{\ell,\sigma}(\epsilon)&=&\sum^{\infty}_{n=-\infty}L_n[f^{<}(\epsilon+n\omega_0){\cal G}^r_{\ell,\sigma}(\epsilon+n\omega_0)\\
\nonumber &+& f^{>}(\epsilon-n\omega_0){\cal
G}^r_{\ell,\sigma}(\epsilon-n\omega_0)],
\end{eqnarray}
where $L_n$ is given by
\begin{equation}
L_{n}(\epsilon)=e^{-g^2(1+N_{B})}e^{\frac{n\omega_0}{2k_BT}}I_n(2g^2\sqrt{N_{B}(1+N_{B})})
\end{equation}
with a boson distribution function of
$N_{B}=1/(e^{\hbar\omega_0/(k_BT)}-1)$, and a Bessel function of
$I_n(x)$. The expressions of $f^{<}(\epsilon+n\omega_0)$ and
$f^{>}(\epsilon-n\omega_0)$ are

\begin{equation}
f^{<}(\epsilon+n\omega_0)=\frac{\Gamma^e_{\ell,L}f_L(\epsilon+n\omega_0)+\Gamma^e_{\ell,R}f_R(\epsilon+n\omega_0)}
{\Gamma^e_{\ell,L}+\Gamma^e_{\ell,R}},
\end{equation}
and
\begin{equation}
f^{>}(\epsilon-n\omega_0)=1-\frac{\Gamma^e_{\ell,L}f_L(\epsilon-n\omega_0)+\Gamma^e_{\ell,R}f_R(\epsilon-n\omega_0)}
{\Gamma^e_{\ell,L}+\Gamma^e_{\ell,R}},
\end{equation}
where the effective tunneling rate of $\Gamma^e_{
\ell,L(R)}=\Gamma_{\ell,L(R)}X^2$ with the reduction factor of
$X=exp[-\frac{1}{2}g^2 coth^2(\hbar\omega_0/(2k_BT))]$. The dressed
electron retarded Green function under $H_{new}$ can be obtained
following the procedure introduced in our previous work.$^{20)}$ We
have the expression of ${\cal G}^r_{\ell,\sigma}$
\begin{equation}
{\cal G}^r_{\ell,\sigma}(\epsilon)=\sum^{8}_{m=1}{\cal
G}^r_{\ell,m,\sigma}(\epsilon)=\sum_{m=1}^{8}\frac{p_m}{\mu_{\ell}-\Sigma_m},
\end{equation}
where $\mu_{\ell}=\epsilon-E^e_{\ell}+i\Gamma^e_{\ell}/2$.
$\Gamma^e_{\ell}=(\Gamma_{\ell,L}+\Gamma_{\ell,R})X^2$. The sum of
$m$ in Eq. (9) has over 8 possible configurations. We consider an
electron (of spin $\sigma$) entering level $\ell$, which can be
either occupied (with probability $N_{\ell,\bar\sigma}$) or empty
(with probability $1-N_{\ell,\bar\sigma}$). For each case, the level
$j$ can be empty (with probability
$a_j=1-N_{j,\sigma}-N_{j,\bar\sigma}+c_j$), singly occupied in a
spin $\bar\sigma$ state (with probability
$b_{j,\bar\sigma}=N_{j,\bar\sigma}-c_j$) or spin $\sigma$ state
(with probability $b_{j,\sigma}=N_{j,\sigma}-c_j$), or a
double-occupied state (with probability $c_j$). Thus, the
probability factors associated with the 8 configurations appearing
in Eq. (9)  become $p_1=(1-N_{\ell,\bar\sigma})a_j$,
$p_2=(1-N_{\ell,\bar\sigma})b_{j,\bar\sigma}$,
$p_3=(1-N_{\ell,\bar\sigma})b_{j,\sigma}$,
$p_4=(1-N_{\ell,\bar\sigma})c_j$, $p_5=N_{\ell,\bar\sigma}a_j$,
$p_6=N_{\ell,\bar\sigma}b_{j,\bar\sigma}$,
$p_7=N_{\ell,\bar\sigma}b_{j,\sigma}$, and
$p_8=N_{\ell,\bar\sigma}c_j$. $\Sigma_m$ in the denominator of Eq.
(9) denotes the self-energy correction due to Coulomb interactions
and coupling with level $j$ in configuration $m$. We have
$\Sigma_1=t_c^{2}/\mu_j$,
$\Sigma_2=U^e_{\ell,j}+t_c^{2}/(\mu_j-U^e_j)$,
$\Sigma_3=U^e_{\ell,j}+t_c^{2}/(\mu_j-U^e_{j,\ell})$,
$\Sigma_4=2U^e_{\ell,j}+t_c^{2}/(\mu_j-U^e_j-U^e_{j,\ell})$,
$\Sigma_5=U^e_{\ell}+t_c^{2}/(\mu_j-U^e_{j,\ell})$,
$\Sigma_6=U^e_{\ell}+U^e_{\ell,j}+t_c^{2}/(\mu_j-U^e_j-U^e_{j,\ell})$,
$\Sigma_7=U^e_{\ell}+U^e_{\ell,j}+t_c^{2}/(\mu_j-2U^e_{j,\ell})$,
and
$\Sigma_8=U^e_{\ell}+2U^e_{\ell,j}+t_c^{2}/(\mu_j-U^e_j-2U^e_{j,\ell})$.
Note that the fractional numbers of the thermally averaged
one-particle occupation number $N_{\ell,\sigma}$ and two-particle
correlation functions $c_{\ell}$ appear only on the probability
weights of ${\cal G}^r_{\ell,m,\sigma}$. Once $t_c=0$ and $g=0$, the
expression of Eq. (9) is the same as Eq. (3) in Ref. 6.
$N_{\ell,\sigma}$ and $c_{\ell}$ can be obtained by solving the
on-site lesser Green's functions,$^{20)}$ their expressions are

\begin{equation}
N_{\ell,\sigma}=-\int \frac{d\epsilon}{\pi}\sum^8_{m=1}
\frac{\Gamma^e_{\ell}f_{\ell}(\epsilon)+\Gamma^{e}_jf_j(\epsilon)}{\Gamma^e_{\ell}+\Gamma^{e}_{j}}
Im{\cal G}^r_{\ell,m,\sigma}(\epsilon),
\end{equation}

and
\begin{equation}c_{\ell}=-\int
\frac{d\epsilon}{\pi}\sum^8_{m=5}
\frac{\Gamma^e_{\ell}f_{\ell}(\epsilon)+\Gamma^{e}_jf_j(\epsilon)}{\Gamma^e_{\ell}+\Gamma^{e}_{j}}
Im{\cal G}^r_{\ell,m,\sigma}(\epsilon).
\end{equation}
Note that $\ell \neq j$ in Eqs. (9)-(11), which are valid in the
condition of $t_c/U_{\ell} \ll 1$. In addition, the limitation of
$0\le N_{\ell,\sigma}(c_{\ell}) \le 1$ should be satisfied.

\section{Results and Discussion}

Bulk metals have very large plasmon frequencies, which can be ten
times larger than the on-site Coulomb interactions of QDs.
Therefore, the effects of EPIs on the tunneling current were ignored
in the previous studies.$^{18-20)}$ The plasmon frequency of
metallic nanostructures is the same order of magnitude as the
on-site Coulomb interactions and the EPIs is strong.$^{13)}$ One can
expect to observe the plasmon assisted tunneling processes in the
tunneling current spectra. To reveal the effects of EPIs on the
tunneling current spectra of QDMs, we initially consider the case
without EPIs. The tunneling current and differential conductance are
plotted in Fig. 2 with $U_{\ell}=U_0=60\Gamma_0$,
$U_{\ell,j}=U_I=20\Gamma_0$, and $E_{\ell}=E_0+\eta_{\ell} eV_a$.
There is a large voltage across the junction, therefore the shift of
energy level $E_{\ell}$ arising from the applied bias is considered
by $\eta_{\ell} eV_a$. We have adopted $\eta_{\ell}=0.5$ based on
the assumption that QDs are located in the central position between
two electrodes. The black lines show the typical staircase structure
of the tunneling current and an oscillatory differential conductance
with respect to the applied bias arising from the intradot Coulomb
interactions. These structures will be washed out with increasing
temperature. We will focus on the transport behavior throughout at
the low temperature of $k_BT=1\Gamma_0$. In the presence of interdot
Coulomb interactions (red lines), new staircase structures appear in
the tunneling current. Five peaks in differential conductance
labeled from $V_1$ to $V_5$ result from electrons of the left
electrode through the resonant channels of $\epsilon_1=E_0$,
$\epsilon_2=E_0+U_I$, $\epsilon_3=E_0+U_0$,
$\epsilon_4=E_0+U_0+U_I$, and $\epsilon_5=E_0+U_0+2U_I$. In the
presence of $t_c$ (see the blue lines), each peaks ($V_1,V_2,V_4$
and $V_5$) split into the bonding (BD) and antiboding (ABD) states.
Such structures can be depicted by using a single molecule with
$E_0-t_c$ and $E_0+t_c$ states filled with one, two, three and four
electrons. The electron filling of such a QDM satisfies Hund's rule.
In addition, peaks $V_2$ and $V_4$ have extra peaks, which
correspond to the spin singlet states (two electrons and three
electrons). For instance the two electron singlet state has a
resonant pole $\epsilon=E_0+U_I-t^2_c/(U_0-U_I)$ resulting from the
$p_2$ in Eq. (9), which is different from the two electron triplet
state with pole $\epsilon=E_0+U_I\pm t_c$ from the $p_3$ of Eq. (9).
The differential conductance structure resulting from three
electrons can also be analyzed from the $p_6$ and $p_7$ of Eq. (9).
On the basis of results in Fig. 2, we find that the interdot Coulomb
interactions play an important role in distinguishing between the
configurations of one electron, two electrons, three electrons and
four electrons. Many theoretical works have been devoted to
investigate the tunneling current through parallel QDs for the
applications of quantum computing.$^{24)}$ Nevertheless, there still
lacks a comprehensive theory to reveal the spin states of parallel
QDs. The retarded Green function of Eq. (9) provides a closed form
expression to distinguish eight configurations in the parallel QDs.

Figure 3 shows the tunneling current (J) and differential
conductance (dG) for different strengths of EPIs at
$U_0=60\Gamma_0$, $U_I=20\Gamma_0$, $t_c=0$,
$\omega_{i,j}=\omega_0=20\Gamma_0$. With increasing the strength of
EPIs, the energy levels of the QDs, intradot Coulomb interactions,
and interdot Coulomb interactions are renormalized by EPIs such as
$E^e_{\ell}=E_{\ell}-g^2\omega_0$, $U^e_{\ell}=U_{\ell}-2g^2
\omega_0$, and $U^e_{\ell,j}=U_{\ell,j}-2g^2\omega_0$. The current
spectra and differential conductance change considerably. Five peaks
of dG labeled from $V_1$ to $V_5$ at g=0 are shifted to the low bias
regime. The magnitude and width of these peaks become smaller and
narrower with increasing $g$. This is attributed to the current
reduction factor of $e^{-g^2(1+N_B)}$ [see Eq. (6)] and reduction of
the tunneling rates $\Gamma^e_{\ell}= \Gamma_{\ell} X^2$. In
addition to these five peaks, there is a satellite peak labeled n=1,
which arises from one plasmon assisted tunneling process. For the
blue line, we see two one-plasmon assisted tunneling peaks
corresponding to $\epsilon=E^e_0+\omega_0$, and
$\epsilon=E^e_0+U^e_I+\omega_0$. As a consequence of the very low
temperatures, the structures arising from $\epsilon=E^e_0-\omega_0$,
and $\epsilon=E^e_0+U^e_I-\omega_0$, which are contributed from the
first term of Eq. (5), correspond to electrons of the QDM with
energy levels $E^e_0$ and $E^e_0+U^e_I$ to have one plasmon emission
process to escape out the QDM. This processes are obviously
suppressed in the small bias regime resulting from small electron
population of the QDM.

Since we consider $t_c=0$, the spin degree of freedom can not be
resolved in Fig. 3.$^{6)}$ To further understand the EPIs on the
current spectra of QDMs with finite $t_c$, we plot the tunneling
current (J) and differential conductance (dG) as a function of the
applied bias for the case of $U_{\ell}=100\Gamma_0$,
$U_{\ell,j}=40\Gamma_0$, and $t_c=6\Gamma_0$ in Fig. 4. The first
two peaks indicated by the black line for dG correspond to the BD
and ABD states. The following three peaks are similar to those
indicated by the blue line in Fig.2. They result from the spin
singlet and triplet states.  For $g=0.5$, these spin-dependent
spectra of dG just shift to the low bias regimes, whereas the
separation between the BD and ABD peaks in the triplet state is not
changed. The exchange energy of the singlet state
$-t^2_c/(U^e_{\ell}-U^e_{\ell,j})$ is also invariant. For $g=1$, the
tunneling currents are enhanced resulting from the absence of the
interdot Coulomb blockade. The spectra from the two particle states
merge into that of one particle at $g=1$. Consequently, the
spin-dependent spectra of dG is suppressed. In addition, we observe
multiple plasmon assisted tunneling processes (n=2).

\section{Summary and conclusions}

In this study we analyzed the effects of homogenous EPIs on the
tunneling current spectra of QDMs in the absence of detected
proteins. As a result of the renormalization of the energy levels of
QDs, intradot and interdot Coulomb interactions, and the tunneling
rates, there is a significant change in the tunneling current
spectra for strong EPIs coupling. Because of the hopping of plasmons
between two metal nanostructures, the indirect interdot plasmon
mediated electron electron Coulomb interactions appear. We predict
that the multiple plasmon assisted tunneling processes can be
observed in the tunneling current spectra of metal core/shell
semiconductor QDs for strong EPIs. In the presence of detected
proteins, which will glue to the QDMs, the DOS of QDMs is changed.
Therefore, the measured tunneling current spectra are tilted to
judge the identity of detected proteins.

\mbox{}\\
{\bf Acknowledgments} This work was supported by the National
Science Council of Taiwan under Contract No: NSC 101-2112-M-
008-014-MY2.

E-mail: mtkuo@ee.ncu.edu.tw\\

\mbox{}\\





\newpage

\begin{figure}[h]
\centering
\includegraphics[scale=0.3]{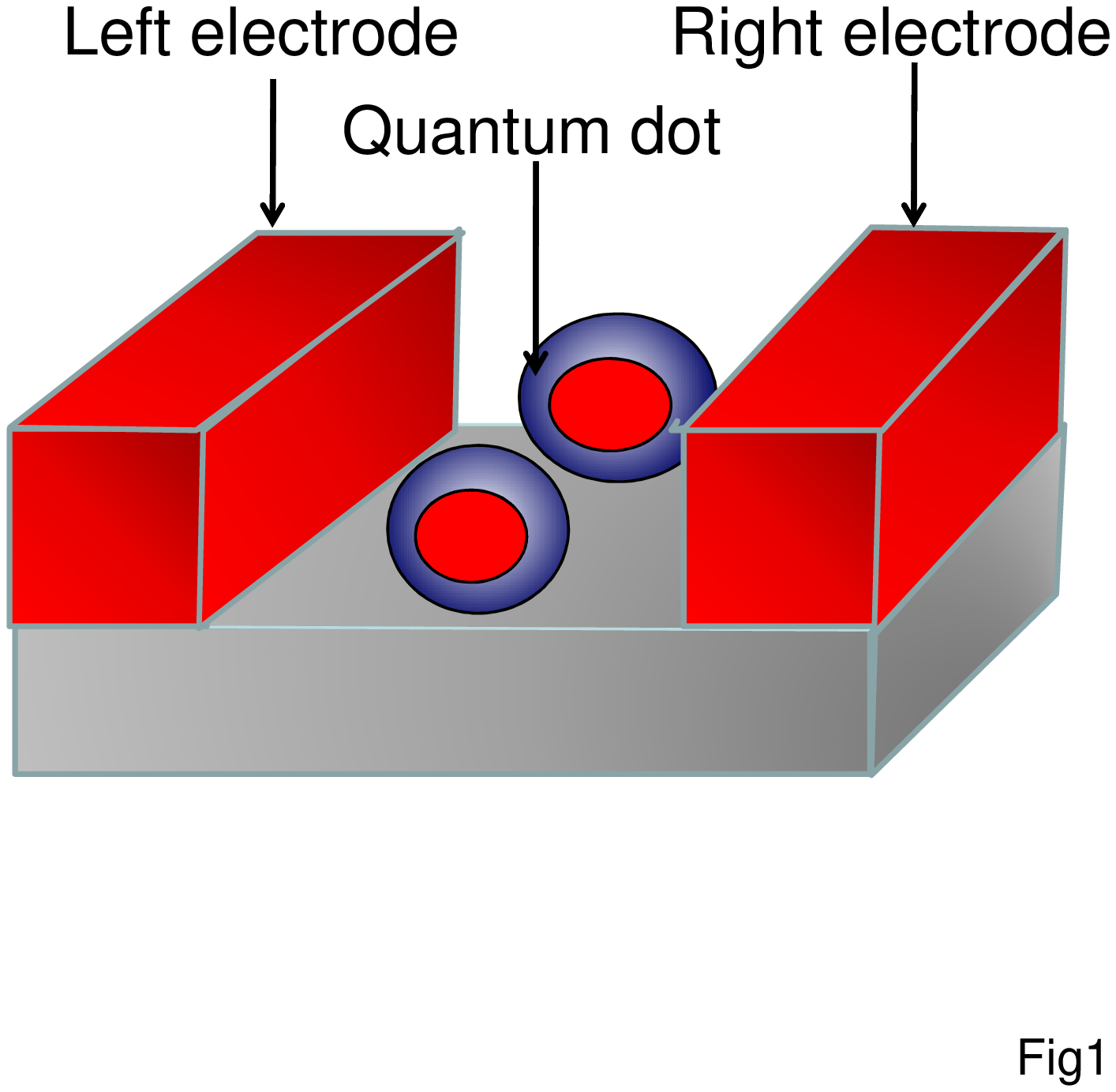}
\caption{A metal core/semiconductor shell quantum dot molecule
embedded in a matrix connected to metallic electrodes.}
\end{figure}

\begin{figure}[h]
\centering
\includegraphics[scale=0.3]{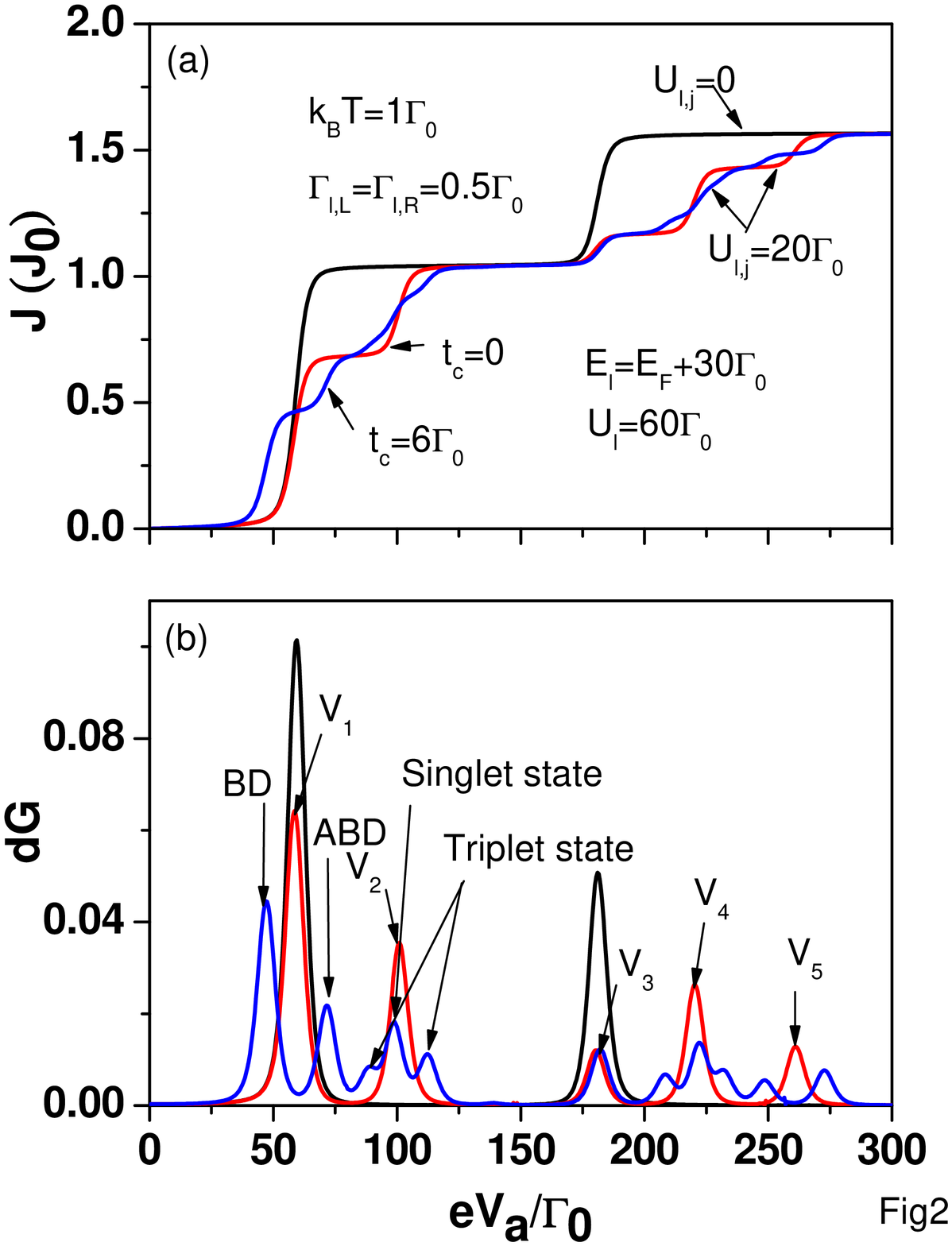}
\caption{(a) Tunneling current J and (b) differential conductance
$dG=dJ/dV_a$ as a function of the applied bias for
$U_{\ell}=U_0=60\Gamma_0$, $E_{\ell}=E_F+30\Gamma_0$,
$k_BT=1\Gamma_0$, and $\Gamma_{\ell,L}=\Gamma_{\ell,R}=0.5\Gamma_0$
in the absence of electron plasmon interactions. Black lines
($U_{\ell,j}=0$), red lines ($U_{\ell,j}=20\Gamma_0$ and $t_c=0$),
and blue lines ($U_{\ell,j}=20\Gamma_0$ and $t_c=6\Gamma_0$). Note
that the tunneling current J is in units of $J_0=e\Gamma_0/h$.}
\end{figure}

\begin{figure}[h]
\centering
\includegraphics[scale=0.3]{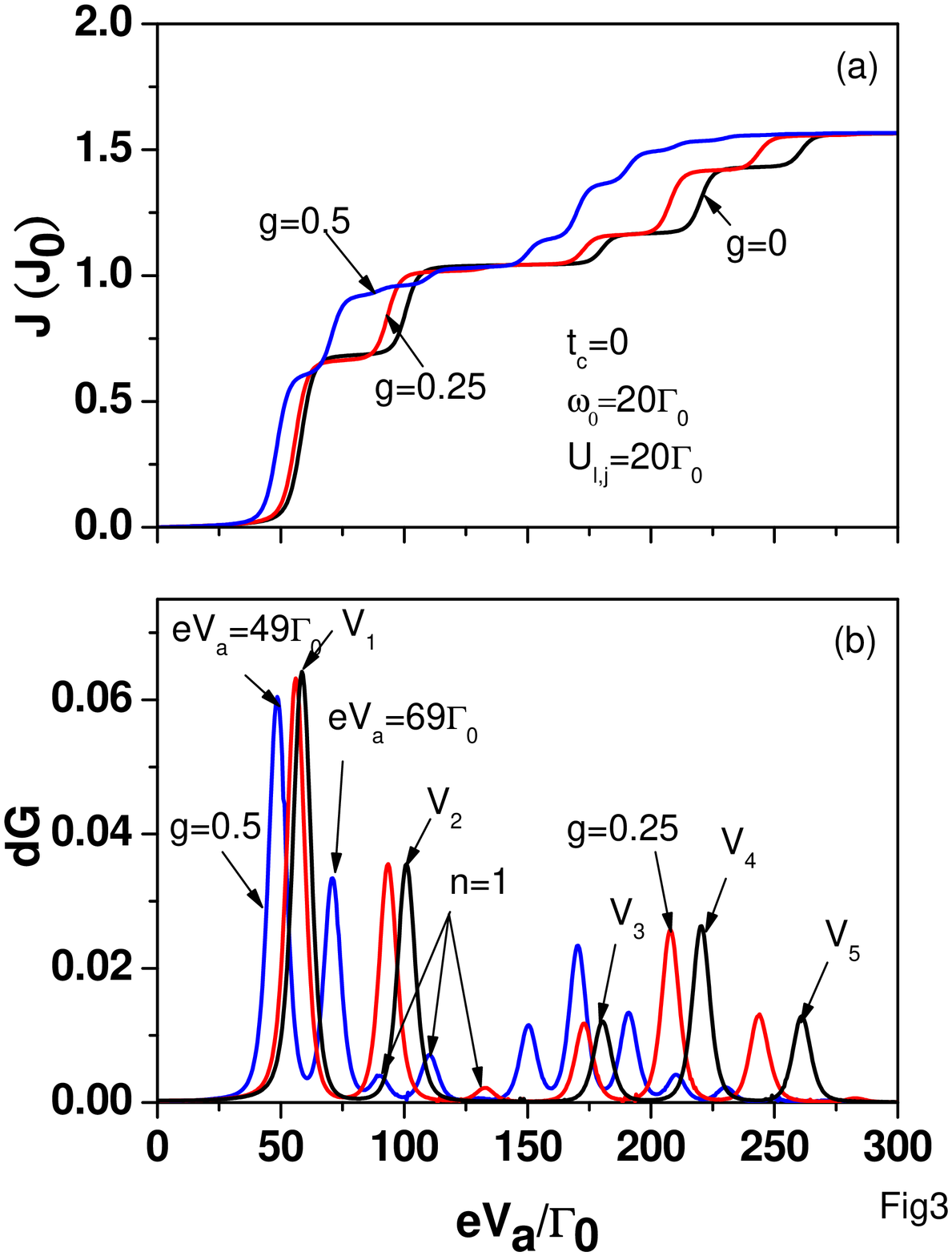}
\caption{(a) Tunneling current J and (b) differential conductance
$dG$ as a function of the applied bias in the presence of electron
plasmon interactions. Black lines ($g=0.0$), red lines ($g=0.25$),
and blue lines ($g=0.5$). We have a plasmon frequency
$\omega_0=20\Gamma_0$. The other physical parameters are the same as
those for the red lines in Fig.2.}
\end{figure}
\begin{figure}[h]
\centering
\includegraphics[scale=0.3]{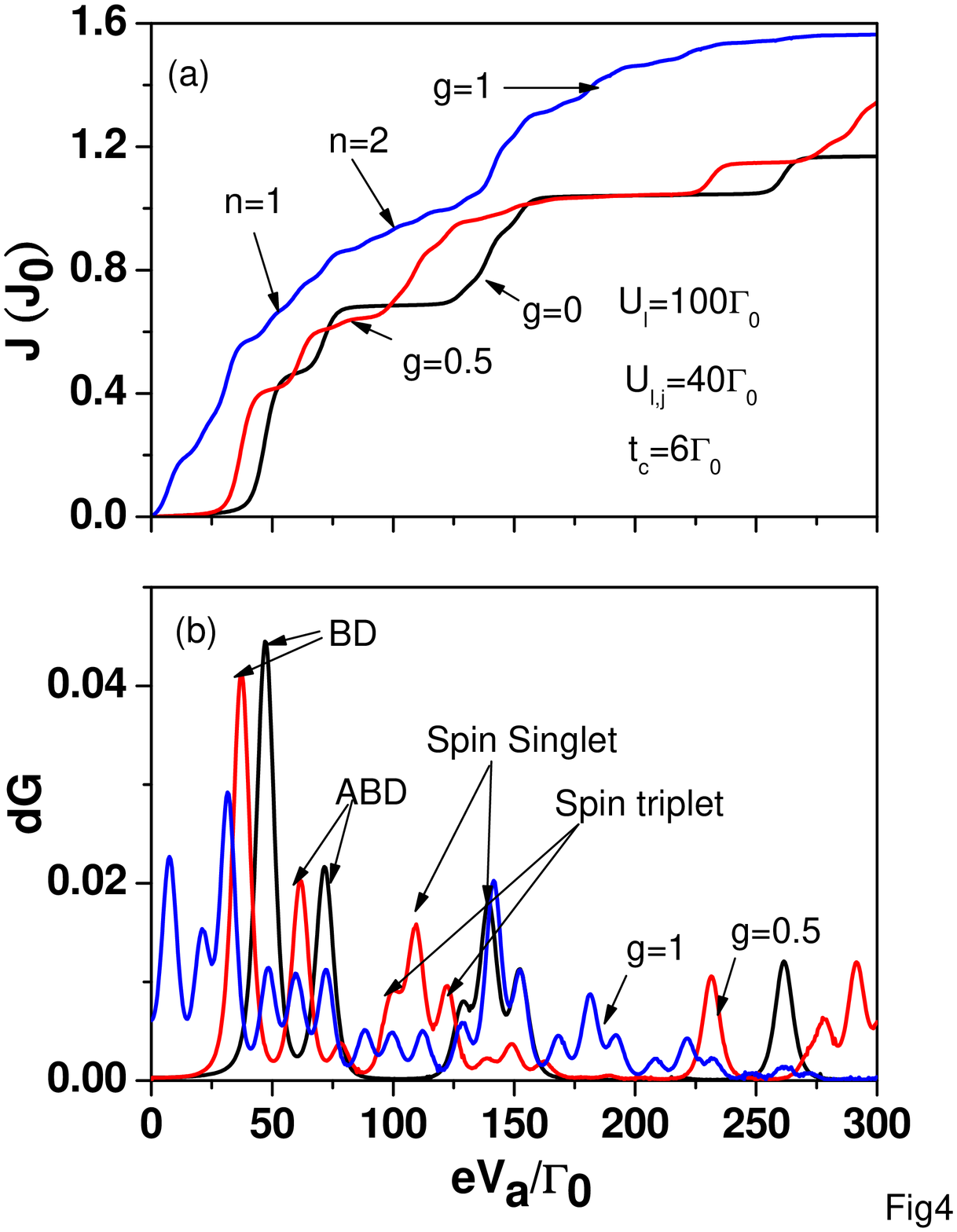}
\caption{(a) Tunneling current J, and (b) differential conductance
$dG$ as a function of the applied bias for different g values. Black
lines ($g=0$), red lines ($g=0.5$), and blue lines ($g=1$). We have
adopted the following physical parameters:$U_{\ell}=100\Gamma_0$,
$U_{\ell,j}=40\Gamma_0$, $t_c=6\Gamma_0$, $k_BT=1\Gamma_0$, and
$\Gamma_{\ell,L}=\Gamma_{\ell,R}=0.5\Gamma_0$.
 }
\end{figure}

\end{document}